\documentclass[preprint,12pt]{elsarticle}

\usepackage{natbib}
\usepackage{amsmath,amssymb}

\begin{document}

\begin{frontmatter}



\title{Fractional diffusion emulates a human mobility network during a simulated disease outbreak}


\author{Kyle B. Gustafson\corref{cor1}\fnref{label1}}
\author{Basil S. Bayati\fnref{label1}}
\author{Philip A. Eckhoff}

\address{Institute for Disease Modeling, Intellectual Ventures, Bellevue, WA, USA}

\cortext[cor1]{kgustafson@intven.com}
\fntext[label1]{Equal contributors}

\begin{abstract}

From footpaths to flight routes, human mobility networks facilitate the spread of communicable diseases. Control and elimination efforts depend on characterizing these networks in terms of connections and flux rates of individuals between contact nodes. In some cases, transport can be parameterized with gravity-type models or approximated by a diffusive random walk. As a alternative, we have isolated intranational commercial air traffic as a case study for the utility of non-diffusive, heavy-tailed transport models. We implemented new stochastic simulations of a prototypical influenza-like infection, focusing on the dense, highly-connected United States air travel network. We show that mobility on this network can be described mainly by a power law, in agreement with previous studies. Remarkably, we find that the global evolution of an outbreak on this network is accurately reproduced by a two-parameter space-fractional diffusion equation, such that those parameters are determined by the air travel network.

\end{abstract}

\end{frontmatter}

\section{Introduction}

Characterization of the spatial spread of a disease is an essential problem for disease control and elimination. Epidemiological models that incorporate spatial dispersion have been used for numerous diseases such as measles \cite{Grenfell2001,Bharti2011}, dengue \cite{Cummings2004,Chao2012,Perkins2013}, seasonal influenza \cite{Viboud2006}, and pandemic flu \cite{Ferguson:2005,Merler2011,VanKerkhove2010}. A recent review of the topic \cite{Riley:2015fl} identified five opportunities for including spatial effects in epidemic models, including the issue of long-distance transport. The underlying spatial process for many epidemiological models is often assumed to be classical diffusion, gravity, or radiation flux. However, these are not necessarily valid for human mobility networks with long-range movements, as was shown for the circulation of US bank notes by Brockmann et al. \cite{Brockmann:2006}.

Their seminal paper used ambivalent fractional-diffusion as a generalized model of transport processes with extreme events, which matched observed trajectories of bank notes. A similar approach, space-fractional diffusion, arises from a generalization of a random walk when the distribution of step sizes in the walk is described by a power law, $x^{-(1+\alpha)}$ with $\alpha\leq2$, rather than a Gaussian function of the step sizes. This type of model has general application  across fields from plasma physics \cite{Gustafson:2012walk} to ecology \cite{Humphries:2010cd,Jacobs:2015gf}. Mathematically, this heavy-tailed distribution of step sizes with an infinite variance produces superdiffusive dispersion, also known as a L\'evy flight. Waiting times between steps are taken to be $\lambda$-scaled within a Poisson process. In practice, a natural process can only be roughly approximated by a L\'evy flight. In some applications, a truncation of the L\'evy flight or a similar L\'evy walk can be a good approximation \cite{Fogleman:2001cc,DelCastilloNegrete:2009dc,Gustafson:2012walk}. Such models  cannot expected to account for all the detailed connections in a non-local network, but may provide a useful and efficient macroscale approximation for a practically heavy-tailed (long-range) distribution of step sizes.

Human displacements occur on many scales through channels ranging from walking to driving to flying. The relative importance of these channels for spatial epidemic models depends on the density and territorial extent of a society, as well as cultural context and socioeconomic development. We decided to focus on the scheduled air traffic network that determines long-range transport in developed countries with large land area, such as the United States, China and India. Clearly, long-range displacements in many contexts will be important for understanding how diseases spread spatially \cite{Peyrard:2008kk}. We took an empirical approach to measuring air travel networks, rather than attempting an analytical result, as other have before \cite{Trapman:2007ij}. These air travel displacements are instrumental in the fast propagation of a disease outbreak across long distances in certain societies. They are a major source of non-diffusive behavior as long flights bestow heavy-tailed distributions of step-sizes, transporting infected individuals long distances effectively instantaneously. 

For the current work, we ran and analyzed spatial simulations of disease dynamics using a flight schedule database to determine passenger transition rates between areas serviced by commercial flights. Each airport acts as a catchment area for a population subject to a susceptible-infected-susceptible (SIS) disease dynamics. We limited our mobility model to air travel since we are interested in highlighting the long-range behavior that is relatively insensitive to local commuting \cite{Balcan22122009, Strano20150651}. Others \cite{Brockmann:2013} have already performed stochastic simulations of a realistic pandemic using global mobility data. Our main intention here was to use the dense air traffic network in the continental United States as an example of a human mobility network that can be approximated with a simple space-fractional diffusion model for transport. Others have recently examined a variety of network structures for disease dynamics \cite{Keeling:2005hk,Leventhal:2015fh}. Additionally, we extended the stochastic fractional-diffusion work presented in \cite{Bayati:2013} with a more accurate, spatially second-order fractional-diffusion method. Our method is related to fractional reaction-diffusion systems previously explored analytically and numerically \cite{CastilloNegrete:2003,Stollenwerk2009,Hanert:2011cm,Hanert:2012ka}. Our results can feasibly be applied by epidemiologists for other transport networks that have heavy tails in the size of displacements between nodes \cite{Draief:2010ct}. 

Our presentation is organized as follows. First we show our analysis of commercial air travel on several countries' national flight networks before focusing on the USA. We extracted a superdiffusive transport exponent from the distribution of step sizes in the network data. We then describe our version of a stochastic SIS simulation to model seasonal influenza, using USA air travel data to form a weighted adjacency matrix connecting populated areas exposed to the disease. We found these results to be in agreement with known seasonal SIS behaviors, validating our new simulations. Next we propose a one-dimensional, space-fractional-diffusion equation as a macroscale substitute for simulating the United States' commercial air traffic mixing network. Remarkably, we found that our space-fractional approximation reproduces the national-average SIS infection curve best when the fractional exponent is similar to the superdiffusive exponent measured from the national flight network. In cases such as the one studied here, where long-range displacements form an important component of mobility, fractional diffusion is a promising macroscale approximation for general measures of epidemics. 

% Results and Discussion can be combined.
\section{Results}
%%%%%%
\subsection{Determining random walk parameters from a flight database}
%%%%%%

\begin{figure}[htp]
  \begin{center}
  	\caption[]{}
 \includegraphics[width= 1\textwidth] {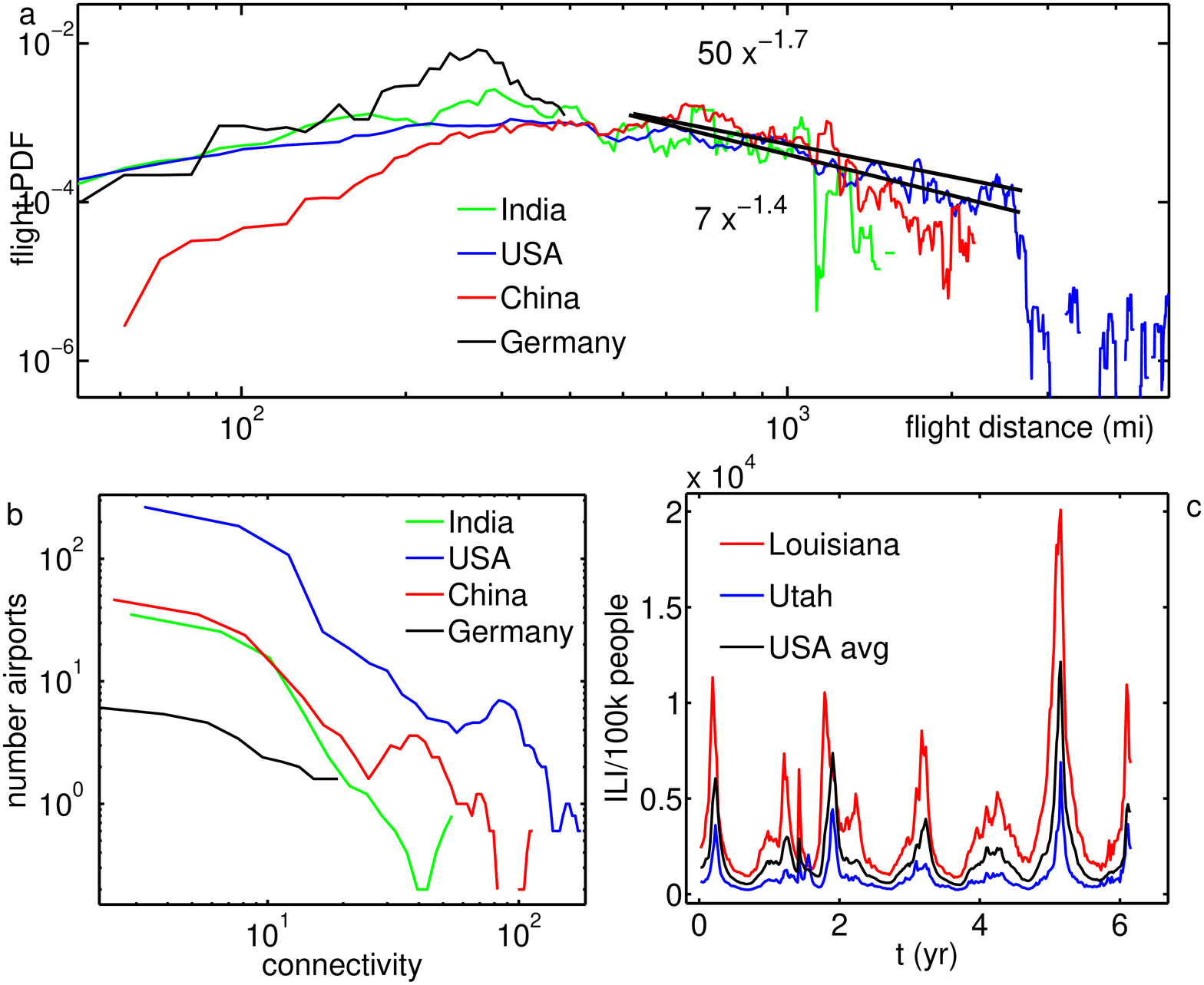}
 \label{fig:figure1} 
  \end{center}
 \end{figure}

In order to approximate a complex human mobility network with a simple random walk model, the statistics of displacements on the network must be determined. We used a commercial air travel database (OAG Aviation Worldwide Limited) to find the probability density of displacements, in miles, for individual travelers. We set the scale of waiting times between steps at one week by aggregating air travel weekly. This should be a reasonable approximation to the travel habits of the average traveler. When aggregating the full set of flight data for several countries' air travel networks, a power law distribution of displacements is readily observed. Countries with large land areas and a dense airport network tend to have a heavy-tailed distribution of air travel step sizes, as shown in Figure \ref{fig:figure1}(a). Here we have accounted for the number of seats in each aircraft and assumed all flights are full. This implies that we are taking an upper bound for mobility by air travel. Other authors have conducted more detailed analyses of air travel networks \cite{Grosche:2007fn}, but our focus was on reproducing large-scale epidemic dynamics with a two-parameter random walk model based on the distribution of step sizes.

Relative to those from the USA and China, displacement distributions from Brazil and India have a shallow tail with a shorter flight range due to the geographic area and distribution of population centers in these countries. The USA network step-size distribution has a superdiffusive power-law exponent $1+\alpha\sim 1.5 \pm 0.2 $ for a significant fraction of displacements in the main part of the tail of the distribution. We fit this power law within the range shown by the black lines using the Nelder-Mead method through the open Matlab toolbox Ezyfit. This power law is truncated to an exponential decay for steps larger than 1500 miles for the USA. We note that a superdiffusive scaling with $\alpha=0.6$ was also inferred from the dispersion of US bank notes \cite{Brockmann:2006}. India, China and the USA show similarities in the shape of the step-size distribution up to the scale of the cutoff. Since we were interested in reproducing this human mobility network with a very simple model, we approximated the US displacement distribution with a superdiffusive power law. For simplicity, we chose to neglect the effects of the naturally occurring truncation that have been discussed for other applications \cite{DelCastilloNegrete:2009dc,Hanert:2014}.

There are two parameters necessary for the partial power law fit indicated in Figure \ref{fig:figure1}(a): $\alpha$ and $D^{\alpha}$ for the slope and intercept of the line in log-log scale. These parameters appear in $q_{\alpha}$, a large step-size ($|x|\to\infty$) limiting solution of a space-fractional diffusion equation (for constant $t$):
\begin{equation}
\label{eq:limitq}
q_{\alpha}(D^{\alpha},x,t) \sim \frac{1}{\pi}\Gamma(\alpha+1)\sin(\alpha\frac{\pi}{2})(D^{\alpha}t)^{1+\frac{1}{\alpha}}|x|^{-(\alpha+1)}.
\end{equation}
Here $\Gamma$ is the extension of the factorial function, $\pi$ is the transcendental number, $t=1$ since our random walk statistics are time-invariant. 
This formula \cite{Mainardi:2007wc} allowed us to estimate the order of magnitude of $D^{\alpha}$, which we found to be $D^{\alpha} \sim 1$ when $\alpha \sim 0.4$ and $D^{\alpha} \sim 10$ when $\alpha \sim 0.7$. This well-grounded, if inexact, calculation of $D^{\alpha}$ is supported by our parameter scan in the final results section. Further discussion of the fractional diffusion equation is given in the Methods.
	
The distributions of airport connectivity for the same four countries are given in Figure \ref{fig:figure1}(b), where connectivity for a given node is simply the number of other nodes with a connecting flight to that node. This plot makes clear that the United States' network is roughly ten times denser than the next example, China, though the shapes of the curves are similar.  With 482 nodes, the US air travel network is significantly denser than any other single nation examined. This network was therefore the focus of our study to find a plausible fractional diffusion approximation for a human mobility network superimposed with disease kinetics.

%%%%%%
\subsection{Basic model for influenza outbreaks on an air travel network}
%%%%%%

As we examined the macroscale dynamics of disease transport, we required a prototypical epidemiological model. We used the susceptible-infected-susceptible (SIS) differential equations (see Methods) with $\beta$ being the rate of infection in units of population per week, and $\gamma$ the rate of recovery in the same units. In this standard model, the basic reproduction number is $R_0=\frac{\beta}{\gamma}\frac{S_0}{N}$, where $S_0$ is the initial number of infected and $N = S+I$ is the total population size, including infected individuals numbering $I$ (see Methods). Since we considered an outbreak at a single location, the fraction $S_0/N$ for the entire network is effectively unity. While our model was sufficiently generalized for any disease with SIS behavior, we made considerations for influenza and influenza-like illnesses (ILI). Since the flu virus evolves from season to season to reinfect populations, we treated the dynamics as effectively SIS. 

\begin{figure}[htp]
  \begin{center}
  	\caption[]{}
 \includegraphics[width= 1\textwidth] {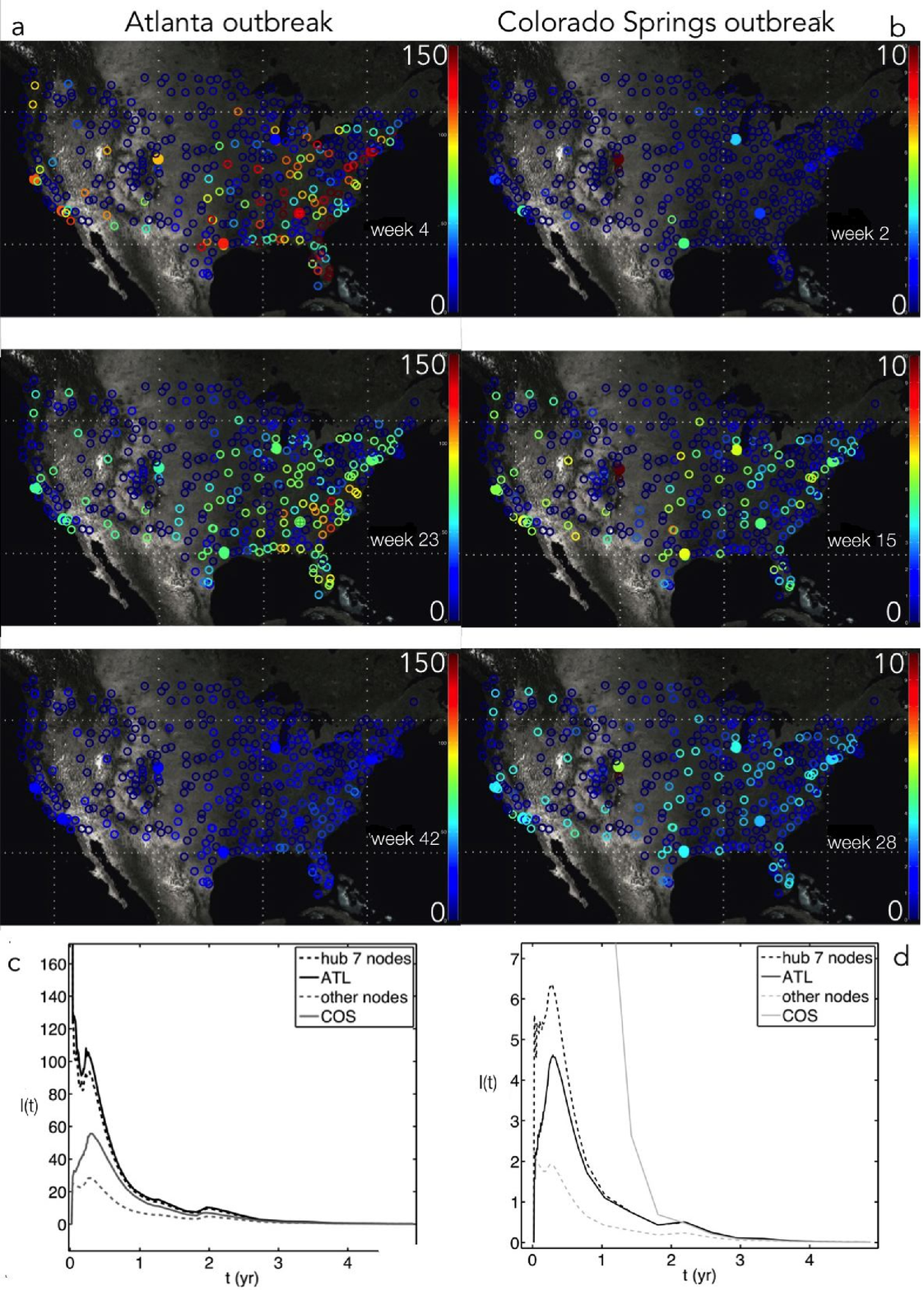}
	 \label{fig:figure2} 
  \end{center}
 \end{figure}

Our air travel network simulations produced node-by-node seasonal SIS disease dynamics initiated by an outbreak of infection at a single airport.  Examples are shown in Figure \ref{fig:figure2} for outbreaks at a highly connected node and a poorly connected node. A highly-connected Atlanta-seeded outbreak with $\beta(t) = \beta_0\mathcal{f}(t)$ where $\mathcal{f}(t)$ is given by historical Google Flu Trends (see Methods and Figure \ref{fig:figure1}(c)) shows a faster and higher amplitude dispersion of the disease compared to a poorly connected Colorado Springs-based outbreak. We chose a $50\%$ initial infected fraction to magnify the properties of the network. The faster growth rate for Atlanta seeding was due to the number of flight connections and the higher peak amplitude of infection is due to the larger population of the Atlanta airport catchment area.  The Atlanta seeded infection quickly moved to the other hubs in the network (the top seven hubs have larger dots) and remained concentrated in the South. 

Due to the reduction of infectivity at seasonal minima and subsequent stochastic fadeout, the outbreak smoothly faded away for both cases within three years for this combination of $\beta=12$ and $\gamma=3$, or $R_0 = 4$.  Again, this model was meant only to provide an example of an SIS outbreak rather than attempting to reproduce the fine structure of influenza dynamics. The expected contrast for a Colorado Springs-seeded outbreak is seen in the right-hand panels (b,d). This outbreak at a less-connected node entered the network at the hubs and traveled more uniformly throughout the country, with a much lower intensity (note difference in scales for $I(t)$). These results indicate that our simulation was functioning as expected. Previously published simulations already noted that the connectivity of a node is significantly more important for spatial disease propagation than physical distance \cite{Brockmann:2013}. Short movies showing the evolution of these outbreaks are provided as electronic supplementary material.

%%%%%%
\subsection{Fractional reaction-diffusion with an outbreak on a line}
%%%%%%

The power law tail for the probability of displacements on the air travel network (Figure \ref{fig:figure1}(a)) can be studied naturally with the non-local space-fractional diffusion operator (see Methods).  In pursuit of a minimal model for non-diffusive transport on a realistic flight traffic network, we restricted the numerical solution of the fractional diffusion equation to one dimension. This avoided the problem of defining the fractional-derivative in two-dimensions and, as shown below, was sufficient for a quantitative agreement, in $\alpha$ and $D^{\alpha}$, for SIS disease dynamics between flight traffic and fractional-diffusion. This agreement was surprising since the one-dimensional fractional diffusion is structured very differently from the air travel network.

We implemented the SIS reaction system into this space-fractional reaction-diffusion model using both deterministic and stochastic solvers for the transport. As demonstrated previously \cite{Bayati:2013}, an operator splitting method with a multinomial kernel can efficiently and accurately solve the space-fractional diffusion equation for a range of $\alpha$ exponent values. These results are shown in Figure \ref{fig:figure3} for constant $\beta = 12$ and $\gamma = 3$. We chose two reference values of $\alpha$ only to highlight the differences between a very superdiffusive $\alpha = 0.3$ and a less superdiffusive $\alpha = 1.3$. The scaling exponent $\eta$ of dispersion with time, $\sigma^2 = D_et^{\eta}$, is more superdiffusive for smaller values of $\alpha$: $\eta = 2/\alpha$. The smaller value of $\alpha = 0.3$ clearly leads to faster spreading of the disease compared to $\alpha = 1.3$. 

For the deterministic case, Figure \ref{fig:figure3}(a,c), we used the solution of the fractional diffusion operator in Equations \ref{eq:dgf1} and \ref{eq:dgf2}. By contrast, we introduced stochastic effects into the fractional-reaction-diffusion system using the Gillespie algorithm as described in the Methods. We used 482 nodes along a line as shown in Figure \ref{fig:figure3}(c-d), seeded at node $i=241$, with 1000 population units per node. The outbreak was seeded with all population units infected at the center node in order to magnify the amplitude and shorten the timescale for analysis. This choice only affects the dynamics quantitatively.

Stochastic behavior is apparent in Figure \ref{fig:figure3}(b,d) that show the temporal and spatial dependence of the fractional SIS model for the same values of $\alpha$ in panels (a) and (c). We saw that stochasticity can affect the time of saturation for $I_i$ for nodes far from the center of the outbreak, though the steady-state mean fraction of infected is the same. The spatial dependence of $I_i$ on the grid position also shows the importance of stochastic reactions for amplifying infection at distal nodes, as well as the appearance of asymmetry due to fluctuations. Without stochastic fluctuations, the time delay to infection saturation is proportional to the distance from the outbreak node. Note that stochastic fluctuations have delayed the saturation time for the outbreak source node, especially in the less superdiffusive ($\alpha = 1.3$/red) case.We chose to plot Figure \ref{fig:figure3}{c} in log-log to highlight the shape of the propagating front, but chose lin-log for panel (d) to highlight the heterogeneity introduced by stochastic effects.

Figure \ref{fig:figure3}(a) shows the time-dependence of $I_i$ for three nodes taken from the symmetric solution shown in Figure \ref{fig:figure3}(c). The number of infected population units decreased for the central outbreak node and rose for other nodes. One of the distal nodes is further from the outbreak, a difference that determined the delay in the arrival of the outbreak, separating the curves in Figure \ref{fig:figure3}(a-b). All nodes eventually reached the same steady state with $I_i(t_{\infty}) \equiv I_{\infty} = \frac{N(\beta-\gamma)}{\beta} = 750$ in the absence of seasonal forcing. Moving away from the center node delayed the arrival, as expected, of the outbreak wave. The shape of the propagating front of the infection in Figure \ref{fig:figure3}(c) was determined partly by the fixed boundary condition, $I_{bnd}(t) = 0$, but this is irrelevant due to the small fraction of population in the distal nodes as $I(t) \to I_{\infty}$. 

In addition to the obvious effect due to the distance from the outbreak node, the time to saturation was also influenced by the fractional diffusion parameters $\alpha$ and $D^{\alpha}$, such that larger (less superdiffusive) values of $\alpha$ imply a longer delay to full network saturation. Following this observation, we exploited the outbreak curve's dependence on the parameters of the fractional diffusion model to find the remarkable comparison described in the next section.

%%%%%%
\subsection{Capturing features of air travel outbreak with fractional diffusion}
%%%%%%

We explored the validity of using a fractional-diffusion model to approximate disease dynamics on the full flight network. For the most straightforward comparison, we excluded stochastic effects and seasonal forcing, since our main conclusions here are not sensitive to these choices. This caused the US air travel epidemic to saturate at $I_{\infty}(\beta,\gamma)$, as we saw for the fractional diffusion model in the previous section. However, while the saturated state was determined by the SIS parameters, the evolution of $I(t)$ was determined partly by the structure of the transport model, as we saw in both the previous sections. Thus, the shape of the network-averaged evolution curve should be a good measure of whether the simplified superdiffusion model can represent the full air travel network. 

We compared the two SIS models using the cumulative difference, $\Theta$ (L2-norm), between $<I(t)>/N$ curves for $0<t<t_{\infty}$. Here, $<I(t)>$ is the network averaged number of infected and $t_{\infty}$ is the time at which the $I_{\infty}$ value is reached. Figure \ref{fig:figure4}(a) shows the direct comparison for a single $R_0$ ($\beta=12$,$\gamma=3$) over a range of fractional derivative $\alpha$ values with $D^{\alpha} = 5$. Better agreement is seen for smaller values of $\alpha$, with $\alpha = 0.5$ providing the best agreement.  Figure \ref{fig:figure4}(b) shows the consistency of the agreement between the two solutions for different sets of $R_0$ for the same values of $\alpha=0.5$ and $D^{\alpha} = 5$.  The asymptotic value $I_{\infty}  = \frac{N(\beta-\gamma)}{\beta}$ depends only on the SIS parameters. The shape of the curve is dependent on both $R_0$ and $\alpha$, as Figure \ref{fig:figure4}(a) shows. 

In Figure \ref{fig:figure4}(c-d), we show $\Theta$ for a parameter scan of $\alpha$ and $D^{\alpha}$ (see Figure \ref{fig:figure1}(a)). The $\Theta$-surface defined by the parameter scan is similar for (c), where $\beta=12$ and $\gamma=3$ and (d) where $\beta=6$ and $\gamma=3$. Recall that $R_0 = \frac{\beta S_0}{\gamma N}$. Regardless of $R_0$, there is a unique minimum in $\alpha$ for each value of $D^{\alpha}$, consistent with the scaling in the long-tail limit of Equation \ref{eq:limitq}. In other words, when $\alpha$ is larger, $D^{\alpha}$ is forced by the form of $q_{\alpha}$ to be larger with a similar scaling as observed in the $\Theta$ scan. While the value of $D^{\alpha}$ is only an estimate, it is clear, at least, that fractional superdiffusion is qualitatively better than classical Brownian diffusion for this human mobility network. 

\section{Discussion}

Devising a campaign strategy for the control and elimination of an infectious disease is a task that can benefit from computational modeling. We sought to encapsulate human mobility-driven spatial spreading of an infectious disease with a fractional reaction-diffusion model as an approximation to the full air travel network. We found good quantitative agreement considering that the mobility network can be characterized only partly by a power law tail. We conclude that a space-fractional diffusion model with superdiffusive statistics can emulate the dynamics of a seasonal disease, such as influenza, on a densely connected, long-range transportation network. This lightweight model offers another tool to explore the possible outcomes of disease outbreaks \emph{in silico}. 

Several caveats apply to the application of the fractional reaction-diffusion approach. In the limit of very few long-range connections, a heavy-tailed process is not a good approximation. Moreover, for a network of very few flights, fractional-reaction-diffusion is inaccurate and unnecessary. We are also aware that our translation of step sizes in miles between airports into a linear series of nodes does not have an obvious mapping function. However, it does seem to produce a compelling agreement. This work could be advanced by considering other modes of transportation such as private vehicles and foot traffic. It may also be useful eventually to consider a 2-D fractional-diffusion model, though the implementation of non-local effects in two dimensions is a challenging area of research \cite{Hanert:2012ka,Hanert:2014}.    

\section{Methods}
\label{sec:Method}

In this work we employed two computational approaches to understand the problem of an oscillatory disease outbreak with spatial dispersion of infectious disease carriers, namely: 1) a stochastic fractional-diffusion-reaction equation and 2) a direct simulation of air traffic transport, the rates of which are derived from real-world flight networks. Both methods use the same algorithm for SIS infection kinetics separately at each node, including an option for stochastic kinetics. Our intent was to find empirically relevant parameter regimes where fractional-reaction-diffusion can reproduce macroscopic features of the spread of an infection on air traffic networks. All simulations were coded in Matlab and each time course (for 482 network nodes in the USA) required no more than 48 hours of serial calculations on a desktop computer. Computational time scaled linearly with the number of nodes as it is dominated by the SIS dynamics.

\subsection{Air travel network simulation}

We built a network reaction-diffusion model in Matlab that allows a great deal of flexibility. Transport can be specified with a matrix of transfer rates from node to node and reactions can be specified with a stochastic simulation algorithm, as described below. Source code is available upon request. For this paper, we included air travel in our simulations using flight schedules taken from the OAG airline database that includes flight data from over 900 airlines. This was the data source for all national flight networks used to build the transfer matrices. For each flight itinerary, the database includes origin and destination airport codes (IATA), number of seats and frequency of flight per week. We computed displacements of passengers summed over a week, assuming that all seats in the airplane are occupied. This approximation is sufficient for tracking infections for diseases with incubation times lasting several days.

Since we were interested in tractable simulations of stochasticity on the network of population centers served by air travel, we approximated each population with a proxy number of individuals, which we call population units. Each node contains this reduced number of units as a metapopulation, in contrast to a full agent-based model simulating each individual. Rather than trying to measure the relative size of each population served by an airport, we used the relative connectivity of each airport service center. Connectivity, $C_n$, for a given node is the number of other nodes with a connecting flight to that node. A minimum node proxy population size is fixed at $N_{min} = 1000$, and the total number of population units at a node, $N_{tot} = N_{min} + B_n C_n$, where $B_n$ scaled the connectivity to the total number of flights between the nodes.  The population units in the simulation could be recalibrated for other applications.

As the population sizes were thus scaled down as described in the previous paragraph, the rate of transport from the flight network was renormalized to give relevant transport rates. In particular, the transfer rate of population was the sum $\sum_i S_i A_i$, where $S_i$ is the number of seats for flight $i$ between the two nodes and $A_i$ is the number of days per week that the connection is serviced. Therefore, the transfer operator was applied on a weekly basis, which sets the time-scale for the simulation. The transfer matrix was also symmetrized to eliminate irregularities in the database in favor of a conservation of total network population. Moreover, the database-derived transfer operator was reduced by a numerical factor ($F = 1\times10^{-7}$) to prevent the artificially small populations from being quickly depleted. Thus, our simulations on the air travel network should be considered as an approximation that preserved the relative size of each population center and the relative transport of population units between them. Applying the transfer operator on a weekly basis assumes that the infected proportion of the population does not change significantly on the weekly timescale. 

\subsection{Fractional diffusion simulation}

A discrete-space, continuous time reaction-diffusion system can be modeled with the master equation, which is a differential equation \cite{Kampen:2007}.  The master equation describes the time evolution of a probability density vector, where each element of the vector denotes a state of the system. In population modeling, this state is typically the number of individuals in certain categories, such as infected with or susceptible to a disease. We are omitting some details of this method that can be found in other work \cite{Bayati:2013}, but including enough here for a degree of completeness.

We let $n_{\lambda}$ denote the number of people at a cell indexed by $\lambda \in \boldsymbol{\lambda}$, $\boldsymbol{n} \triangleq \{ n_{\lambda} \}_{\lambda \in \boldsymbol{\lambda}}$, and $p = p(\boldsymbol{n}, t)$ the probability of finding $n_{\lambda}$ individuals at time $t$.  Then, the master equation is \cite{Kampen:2007}:
 \begin{eqnarray}
\label{eq:Mequation}
\frac{\partial p}{\partial t} &=&  \mathfrak{A}p + \mathfrak{B}p,
\end{eqnarray}
where the operator $\mathfrak{A}$ denotes diffusion, $ \mathfrak{B}$ denotes the transitions due to reactions. The master equation has an initial condition specified by $p(\boldsymbol{n}, 0) = p_0(\boldsymbol{n})$, where $ p_0(\boldsymbol{n})$ is typically a delta function. The formal solution to the master equation is:
\begin{eqnarray}
\label{eq:MequationSolution}
p(\boldsymbol{n}, t) &=& e^{t(\mathfrak{A}+\mathfrak{B})} p(\boldsymbol{n}, 0). 
\end{eqnarray}
\label{subsec:OperatorSplitting}
As discussed in \cite{Bayati:2011} the master equation in some instances can be solved analytically, and therefore it is beneficial to partition the master equation and use analytical solutions when possible.  To this end, we split the operators in the reaction-diffusion equation in order to use analytical solutions for the diffusive part, which yields \cite{Jahnke:2010}:
%\begin{eqnarray}
%\label{eq:MequationSolution}
%|e^{\Delta t \mathfrak{A}} e^{\Delta t  \mathfrak{B} } - e^{\Delta t(\mathfrak{A}+\mathfrak{B})}| = C_1 \Delta t^2, 
%\end{eqnarray}
%where $C_1$ is a constant, and we can therefore write: 
\begin{eqnarray}
\label{eq:MequationSolutionSplit}
p(\boldsymbol{n}, \Delta t) &=&e^{\Delta t \mathfrak{A}} e^{\Delta t  \mathfrak{B}} p(\boldsymbol{n}, 0) + \mathcal{O}(\Delta t^2), 
\end{eqnarray}
which is equivalent to a simulation method where integration in time can be performed via the composition of the two operators:
\begin{eqnarray}
\label{eq:operatorSplitting}
\boldsymbol{\tilde{n}}^{(s)}(t+\Delta t) = \left( \boldsymbol{\phi}_{\mathfrak{A}}^{(\Delta t)} \circ \boldsymbol{\phi}_{\mathfrak{B}}^{(\Delta t)} \right) \left( \boldsymbol{\tilde{n}}^{(s)}(t) \right), 
\end{eqnarray}
for independent samples $s = 1, \ldots, N$.  In other words, the diffusion process is simulated over a time-step $\Delta t$ and the system is updated, then the reactions are simulated with another time-step $\Delta t$.  These two steps constitute a single iteration in the numerical method.  

The reactions operator is discussed below.  The diffusion operator $\boldsymbol{\phi}_{\mathfrak{A}}^{(\Delta t)}$ consists of  transitions that are not necessarily limited to adjacent cells. 
% Jahnke and Huisinga in \cite{Jahnke:2007} proved that the solution to a set of unimolecular transitions is a multinomial distribution.   Let $n_{\lambda'}$ people be at position $\lambda'$ at time $t=0$ and $p(\boldsymbol{n}, 0) = \delta_{\lambda, \lambda'}$, and consider $\mathfrak{B}\equiv 0$, i.e. the diffusion operator in isolation, then  \cite{Jahnke:2007}:
%\begin{eqnarray}
%p(\boldsymbol{n}, \Delta t) = e^{\Delta t \mathfrak{A}} p(\boldsymbol{n}, 0) = \mathcal{M}\left( \lambda, n_{\lambda'},  \mathfrak{G}^{(\lambda')}(\Delta t) \right), 
%\end{eqnarray}
%where $\mathcal{M}\left( \lambda, n_{\lambda},  \mathfrak{G}^{(\lambda')}(\Delta t) \right)$ is  a multinomial distribution and $ \mathfrak{G}^{(\lambda')}(\Delta t)$ is a probability density vector.  
We define the fractional-diffusion kernel using a spatially second-order method (see Appendix, where the discretization is described in detail):
\begin{eqnarray}
\label{eq:dgf1}
\mathfrak{G}^{(\lambda')}_{\lambda'}( \Delta t) &=& 1-\frac{D \Delta t}{h^{\alpha}} \frac{\Gamma(\alpha+1)}{ \Gamma \left( \frac{\alpha}{2}+1 \right)^2} \\ 
\label{eq:dgf2}
\mathfrak{G}^{(\lambda')}_{\lambda' \pm j}( \Delta t) &=&\frac{D \Delta t}{h^{\alpha}}  \frac{(-1)^{j+1} \Gamma(\alpha+1)}{\Gamma \left( \frac{\alpha}{2}-j+1 \right) \Gamma \left( \frac{\alpha}{2}+j+1 \right)}, ~~|j| \ge 1
\end{eqnarray}
where $\Gamma(.)$ denotes the $\Gamma$ function, $D$ denotes the diffusion coefficient, $\mathfrak{G}^{(\lambda')}_j$ denotes the $j$th component of the vector $\mathfrak{G}^{(\lambda')}$, $h$ the size of the compartment, $\alpha$ the value of the fractional derivative, and $\sum_{j}^{} \mathfrak{G}_j(t) = 1$ 
provided that the following time-step criterion is satisfied:
\begin{eqnarray}
\label{eq:timestepmax}
\Delta t &\leq& \frac{h^{\alpha}}{D^{\alpha}} \frac{\Gamma \left( \frac{\alpha}{2} + 1 \right)^2}{\Gamma(\alpha+1)}. 
\end{eqnarray}
In practice, the timestep should be chosen as an order of magnitude less than the right-hand side of Equation \ref{eq:timestepmax}. The spatial boundary condition fixes the value of the population at zero.
 If $\mathcal{U} = \mathcal{U}(x,t)$ denotes the concentration of people at position $x \in \mathbb{R}$ at time $t$, then the expected value of the method described above converges to the symmetric, space-fractional diffusion equation \cite{Mainardi:1997,Gorenflo:2002,Gorenflo:2002b,Gorenflo:1998,Bayati:2011}: 
 \begin{eqnarray}
\label{eq:fracDiffusion}
\frac{\partial \mathcal{U}}{\partial t} &=& D^{\alpha} \frac{\partial^{\alpha}\mathcal{U}}{\partial |x|^{\alpha}}, 
\end{eqnarray}
where $\alpha \in (0,2]$ and $D^{\alpha}$ is the effective diffusion coefficient.

\subsection{Reaction Dynamics}

There has recently been significant work on efficiently simulating spatial processes using the stochastic simulation algorithm (SSA) or Gillespie's algorithm \cite{Gillespie:1976,Gillespie:1977}.  In these models, the domain is discretized into cells such that each cell is assumed to be homogeneous.  The result of the discretization is a set of random variables that denote the number of discrete quantities in each cell  \cite{Elf:2004}.  Diffusion events are modeled as transitions to adjacent cells, where the rate incorporates the macroscopic diffusion coefficient and the distance between the cells.  The transition rates are derived by using finite-volume methods and the spatial process has been shown to converge to classical diffusion in the limit of a large numbers \cite{Bernstein:2005,Bayati:2011}.  Stochastic simulation algorithms can be used to simulate the reactions that occur within cells due to the homogeneity assumption.  Recently there has been a marked interest in developing efficient algorithms for the simulation of inhomogeneous systems with classical diffusion processes  \cite{Lampoudi:2009,Drawert:2010,Bayati:2011,Elf:2004,Hellander:2007,Hellander:2012}.   

For our simulations of fractional diffusion and air traffic, each node has a finite population subject to infectious disease kinetics. We use the standard susceptible-infected-recovered (SIR) and susceptible-infected-susceptible (SIS) models that we implemented numerically with either a deterministic or a stochastic simulation. Conceptually, the infection kinetics are applied to the nodes with a reaction operator $\boldsymbol{\phi}_{\mathfrak{B}}^{(\Delta t)}$. Since we are generally interested in studying stochastic effects, such as fade-out and bursting, we considered various stochastic simulation algorithms, such as Gillespie's algorithm \cite{Gillespie:1976,Gillespie:1977}, $\tau$-leaping \cite{Gillespie:2001}, or $R$-leaping \cite{Auger:2006} (see \cite{Gillespie:2007} for a review).  For an overview of the master equation formulation of stochastic kinetics, see Van Kampen \cite{Kampen:2007}. 
We successfully tested the code with the SIR model system, we focus here on the standard SIS system, where the infection contact rate potentially varies with time:

\begin{align}
\frac{\text{d}S}{\text{d}t}& = -\beta(t) S I / N + \gamma I \\
\frac{\text{d}I}{\text{d}t}& = \beta(t) S I / N - \gamma I
\end{align}
where $I$ is the number of infected, $S$ is the number of susceptible, $N$ is the total population and $\beta$ and $\gamma$ are parameters that determine the reproduction number $R_0=\frac{\beta}{\gamma}\frac{S_0}{N}$ .

The $\beta$ and $\gamma$ parameters affect the rate at which the infection grows and the steady-state infected fraction. The variation in infection rate can be modeled with a simple sinusoidal function, or as we have implemented, with a time-course explicity fit to the Google Flu Trends (GFT) database \cite{Ginsberg2009}. The GFT data gives an estimate of the occurrence of influenza-like illness (ILI) according to internet searches. While the predictive power of GFT data has recently been challenged \cite{Lazer14032014}, the qualitative pattern of the GFT data is descriptive of influenza occurrence. In simulations of the USA flight network, we use the historical GFT time course for individual states, shown in Figure \ref{fig:figure1}(c). 

We scanned through $\gamma$ and $\beta$ to determine whether it is possible to use the fractional-reaction diffusion equation to minimally model the epidemic on the air travel network. We note that $R_0$ in pandemics such as the 1918 flu has been estimated to be in the range of range $1.4$ to $2.8$ \cite{Coburn2009}. Simulations with values of $R_0$ closer to unity exhibit greater stochastic effects \cite{Bayati2012}.  In our results, a value of $R_0=4$ or $R_0=2$ is used as a reference scenario without having qualitative influence on the result.     

\section{Acknowledgments}
The authors thank Bill and Melinda Gates for their active support of this work and their sponsorship through the Global Good Fund. Thanks also to Alexandre Bovet, Daniel Klein, Edward Wenger and Josh Proctor for stimulating conversations.

\section{Author Contributions}
BSB and PAE conceived the project.  KBG analyzed the flight data and performed the numerical simulations and applications.  BSB derived and tested the numerical method.  BSB, KG, and PAE wrote the manuscript.   

\section{Competing interest}

We declare no competing interests with respect to this work.

\clearpage

\section{Figure and movie captions} 

\subsection{Figure 1}
(a) Shows probability of a flight/step of a certain number of miles within a week-long time interval for displacements on the air travel networks in the USA, China, Brazil and India. Considering the partial power law scaling $-(1+\alpha)$ of this data, black lines are fit for $\alpha=0.3$, $\alpha=0.6$, $\alpha=0.9$; (b) Distribution of connectivity for these four countries flight networks; (c) Distribution of single flight displacements for the top seven most densely connected nodes on the US network, showing a deviation from Gaussian with a heavy tail; (d) Google Flu Trends for the USA average and for two states with noticeably different histories.

\subsection{Figure 2}
(a) and (b) Outbreak of a stochastic seasonal SIS influenza-like infection on the USA flight network, with origins either in Atlanta (ATL) (a) or Colorado Springs (COS) (b). The seasonal force of infection is scaled using the Google Flu Trends timeseries for infectivity $\beta(t)$, where the baseline $\beta_0=12$ and $\gamma=3$. After some time (top to bottom), the disease propagates throughout the network and tends to fadeout due to stochastic fluctuations and dilution. The colorbar shows the number of infected population units for each node, $I(t)$. (c-d) Time course of SIS outbreaks for the networks in the above panels, showing the average infected for the seven most connected hubs, the average infected for all other nodes, and the specific nodes ATL and COS. 

\subsection{Figure 3}
One dimensional spatial network SIS simulations with $R_0=4$ using fractional diffusion. One thousand population units are simulated at each node  Blue denotes a very superdiffusive $\alpha = 0.3$ and red denotes a less superdiffusive $\alpha = 1.3$ in all panels. (a) Time dependence of the number of infected $I_i(t)$, for three nodes on the grid for two different values of $\alpha$.  Solid lines denote the node where the infection was seeded. The dashed lines show data from non-central nodes, where the outbreak arrives later for the node further from the center. (b) Same as previous except with stochastic fluctuations. (c) Spatial dependence of the infection for three time-points, denoted by $t_1,t_2,t_3$ for both values of $\alpha$ without stochastic effects. Several curves are shown to demonstrate that, as time progresses, the infection rises at distal nodes. (d) Same as panel (c) except with stochastic dynamics, showing left/right asymmetry and long range fluctuations. The early timesteps are excluded in order to emphasize the fluctuations during a single simulation.

\subsection{Figure 4}
(a) Time traces for a constant $R_0=4$, SIS-modeled outbreak.  The USA flight network (solid curve) is compared to discretized fractional diffusion (dashed lines). As the fractional diffusion exponent $\alpha$ decreases the transport becomes more superdiffusive and the dashed line is closer to the USA network result. The best-fit value of this exponent, $\alpha = 0.5$, is similar to the one measured in Figure \ref{fig:figure1}(a). Smaller values of $\alpha$ cause the curve to overshoot the air travel result. (b) Comparison between fractional diffusion and USA flight network outbreaks for superdiffusive $\alpha=0.5$ for several values of $R_0$ using SIS dynamics at each node. (c-d) Result of parameter scan through $\alpha$ and $D^{\alpha}$ displaying the absolute distance $\Theta$ between the growth curves for average number of infected. The two panels show the difference for constant values of (c) $R_0=2$ and (d) $R_0=4$. These scans show that there is an optimal value of $\alpha$ that nearly matches the air travel model.

\subsection{Movie 1}
Outbreak of a seasonal, stochastic SIS infection in the Atlanta International airport (ATL) catchment area. The colorbar represents the number of infected population units at each node of the air travel network. Population sizes are scaled relatively to the size of the catchment area. The infection spreads quickly to hubs in the network and then fades out as the seasonal infectivity falls and stochastic fluctuations eliminate infected units. The average number of infected for this simulation is shown in Figure \ref{fig:figure3}(c).

\subsection{Movie 2}
\begin{figure}[htp]
  \begin{center}
  	\caption[]{}
 \includegraphics[width= 1\textwidth] {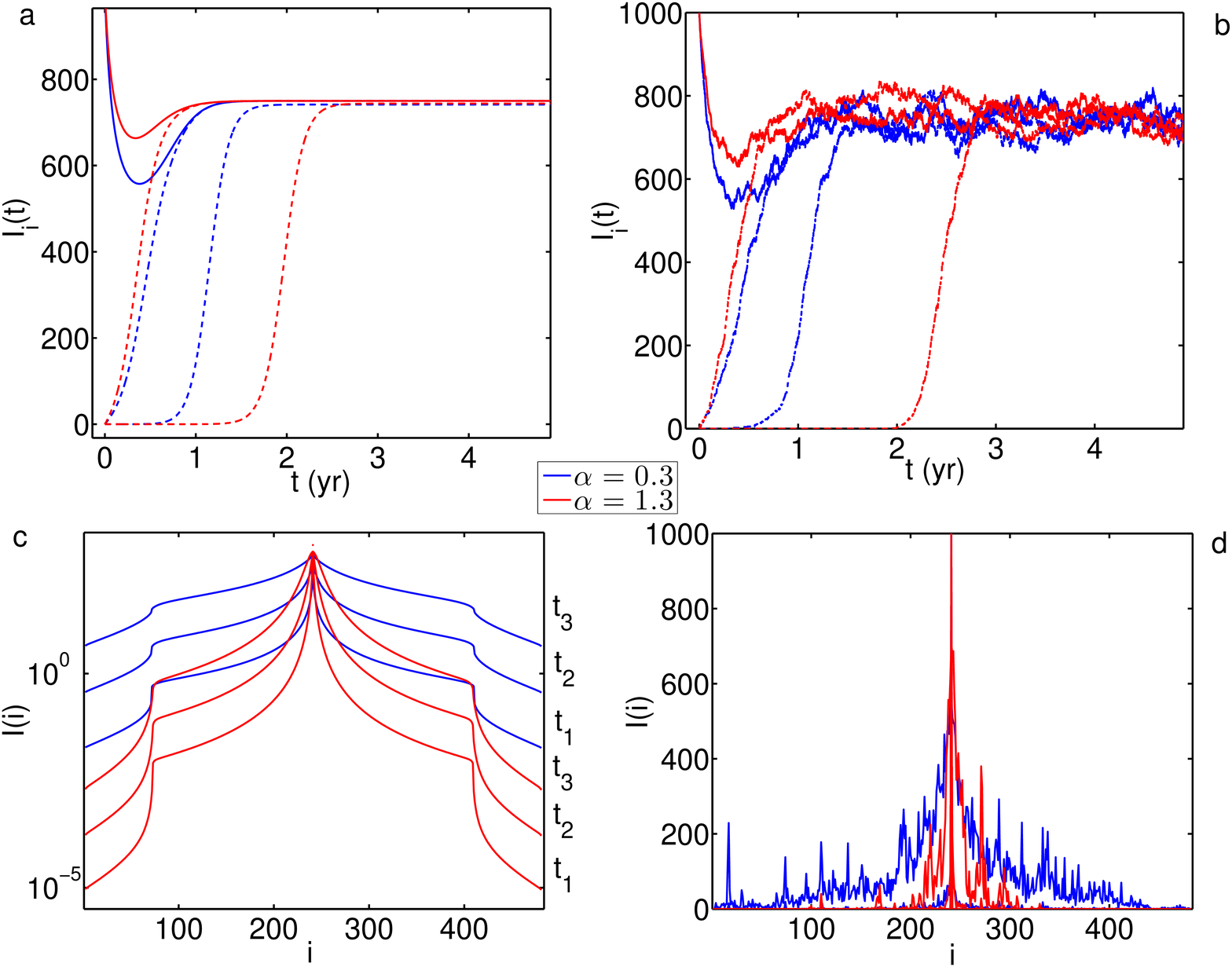}
	 \label{fig:figure3} 
  \end{center}
 \end{figure}

\begin{figure}[htp]
  \begin{center}
  	\caption[]{}
 \includegraphics[width= 1\textwidth] {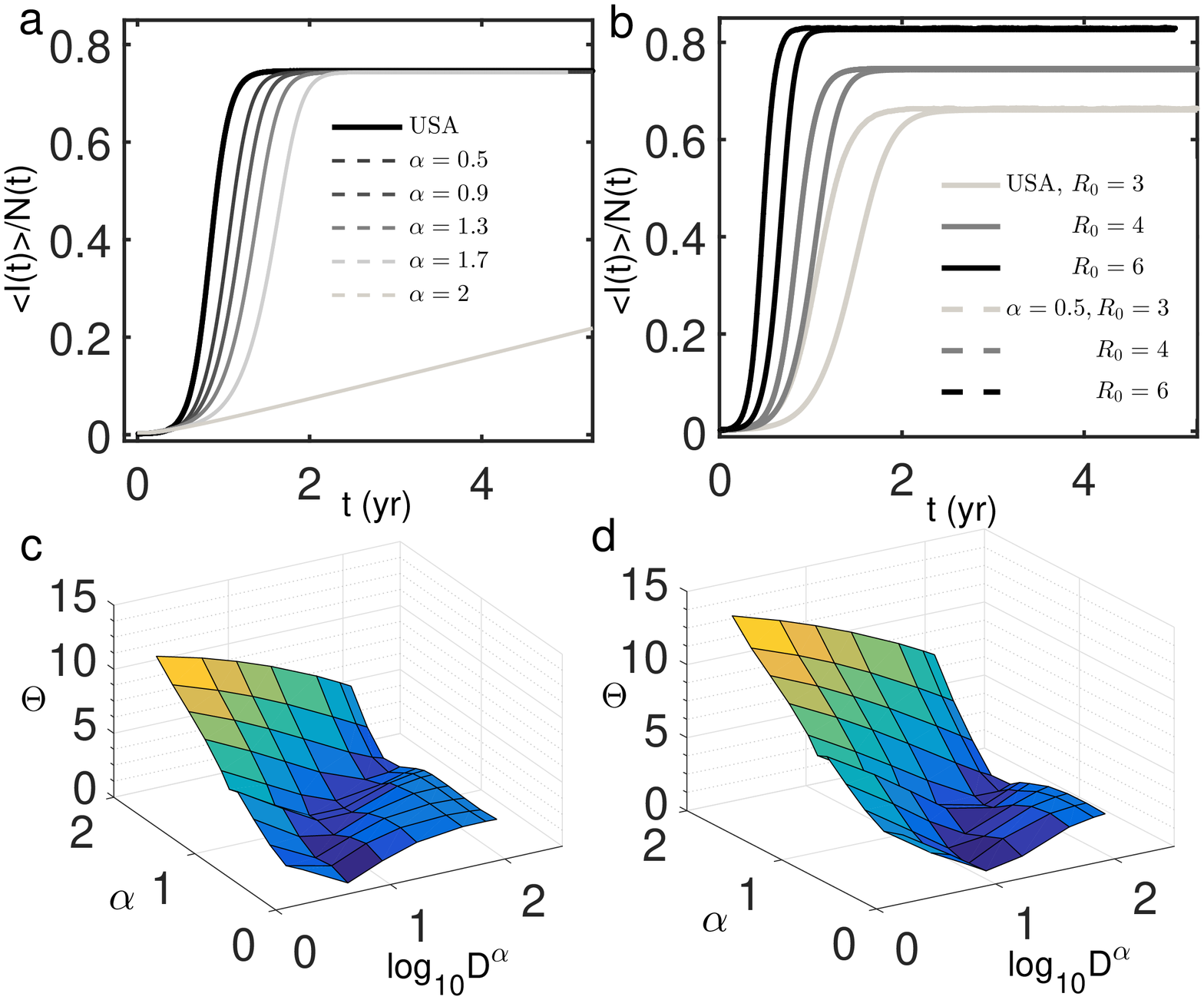}
	 \label{fig:figure4} 
  \end{center}
 \end{figure}

Outbreak of a seasonal, stochastic SIS infection in the Colorado Springs airport (COS) catchment area. The colorbar represents the number of infected population units at each node of the air travel network. Population sizes are scaled relatively to the size of the catchment area. The infection spreads quickly to hubs in the network and then fades out as the seasonal infectivity falls and stochastic fluctuations eliminate infected units. The average number of infected for this simulation is shown in Figure \ref{fig:figure3}(d).

\clearpage

 %%%%						%%%%
\clearpage
%%%%						%%%%
 %%%%							%%%%
 \clearpage
%%%%						%%%%
 %%%%						%%%%
 \clearpage
%%%%						%%%%

\clearpage
%
%\bibliographystyle{kp}
%
%\bibliography{References}

\begin{thebibliography}{54}
	\expandafter\ifx\csname natexlab\endcsname\relax\def\natexlab#1{#1}\fi
	
	\bibitem[Grenfell et~al.(2001)Grenfell, Bjornstad, and Kappey]{Grenfell2001}
	B.~T. Grenfell, O.~N. Bjornstad, and J.~Kappey, ``Travelling waves and spatial
	hierarchies in measles epidemics'', {\em Nature} {\bfseries 414} (2001),
	no.~6865, 716--723.
	
	\bibitem[Bharti et~al.(2011)Bharti, Tatem, Ferrari, Grais, Djibo, and
	Grenfell]{Bharti2011}
	N.~Bharti, A.~J. Tatem, M.~J. Ferrari, R.~F. Grais, A.~Djibo, and B.~T.
	Grenfell, ``Explaining seasonal fluctuations of measles in niger using
	nighttime lights imagery'', {\em Science (New York, N.Y.)} {\bfseries 334}
	(2011), no.~6061, 1424--1427.
	
	\bibitem[Cummings et~al.(2004)Cummings, Irizarry, Huang, Endy, Nisalak,
	Ungchusak, and Burke]{Cummings2004}
	D.~A. Cummings, R.~A. Irizarry, N.~E. Huang, T.~P. Endy, A.~Nisalak,
	K.~Ungchusak, and D.~S. Burke, ``Travelling waves in the occurrence of dengue
	haemorrhagic fever in thailand'', {\em Nature} {\bfseries 427} (2004),
	no.~6972, 344--347.
	
	\bibitem[Chao et~al.(2012)Chao, Halstead, Halloran, and Longini]{Chao2012}
	D.~L. Chao, S.~B. Halstead, M.~E. Halloran, and J.~Longini, Ira~M.,
	``Controlling dengue with vaccines in thailand'', {\em PLoS Negl Trop Dis}
	{\bfseries 6} (2012), no.~10, e1876--.
	
	\bibitem[Perkins et~al.(2013)Perkins, Scott, Le~Menach, and Smith]{Perkins2013}
	T.~A. Perkins, T.~W. Scott, A.~Le~Menach, and D.~L. Smith, ``Heterogeneity,
	mixing, and the spatial scales of mosquito-borne pathogen transmission'',
	{\em PLoS Comput Biol} {\bfseries 9} (2013), no.~12, e1003327.
	
	\bibitem[Viboud et~al.(2006)Viboud, Bjørnstad, Smith, Simonsen, Miller, and
	Grenfell]{Viboud2006}
	C.~Viboud, O.~N. Bjørnstad, D.~L. Smith, L.~Simonsen, M.~A. Miller, and B.~T.
	Grenfell, ``Synchrony, waves, and spatial hierarchies in the spread of
	influenza'', {\em Science} {\bfseries 312} (2006), no.~5772, 447--451.
	
	\bibitem[Ferguson et~al.(2005)Ferguson, Cummings, Cauchemez, Fraser, Riley,
	Meeyai, Iamsirithaworn, and Burke]{Ferguson:2005}
	N.~M. Ferguson, D.~A. Cummings, S.~Cauchemez, C.~Fraser, S.~Riley, A.~Meeyai,
	S.~Iamsirithaworn, and D.~S. Burke, ``Strategies for containing an emerging
	influenza pandemic in southeast asia.'', {\em Nature} {\bfseries 437} (2005),
	no.~39, 209--214.
	
	\bibitem[Merler et~al.(2011)Merler, Ajelli, Pugliese, and Ferguson]{Merler2011}
	S.~Merler, M.~Ajelli, A.~Pugliese, and N.~M. Ferguson, ``Determinants of the
	spatiotemporal dynamics of the 2009 h1n1 pandemic in europe: Implications for
	real-time modelling'', {\em PLoS Comput Biol} {\bfseries 7} (2011), no.~9,
	e1002205--.
	
	\bibitem[Van~Kerkhove et~al.(2010)Van~Kerkhove, Asikainen, Becker, Bjorge,
	Desenclos, dos Santos, Fraser, Leung, Lipsitch, Longini, McBryde, Roth, Shay,
	Smith, Wallinga, White, Ferguson, Riley, and for the WHO Informal Network for
	Mathematical Modelling for Pandemic Influenza H1N1 2009 (Working Group~on
	Data~Needs)]{VanKerkhove2010}
	M.~D. Van~Kerkhove, T.~Asikainen, N.~G. Becker, S.~Bjorge, J.-C. Desenclos,
	T.~dos Santos, C.~Fraser, G.~M. Leung, M.~Lipsitch, J.~Longini, Ira~M., E.~S.
	McBryde, C.~E. Roth, D.~K. Shay, D.~J. Smith, J.~Wallinga, P.~J. White, N.~M.
	Ferguson, S.~Riley, and for the WHO Informal Network for Mathematical
	Modelling for Pandemic Influenza H1N1 2009 (Working Group~on Data~Needs),
	``Studies needed to address public health challenges of the 2009 h1n1
	influenza pandemic: Insights from modeling'', {\em PLoS Med} {\bfseries 7}
	(2010), no.~6, e1000275.
	
	\bibitem[Riley et~al.(2015)Riley, Eames, Isham, Mollison, and
	Trapman]{Riley:2015fl}
	S.~Riley, K.~Eames, V.~Isham, D.~Mollison, and P.~Trapman, ``{Five challenges
		for spatial epidemic models}'', {\em Epidemics} {\bfseries 10} (2015) 68--71.
	
	\bibitem[Brockmann et~al.(2006)Brockmann, Hufnagel, and Geisel]{Brockmann:2006}
	D.~Brockmann, L.~Hufnagel, and T.~Geisel, ``The scaling laws of human travel'',
	{\em Nature Letters} {\bfseries 439} (2006) 462--465.
	
	\bibitem[Gustafson and Ricci(2012)]{Gustafson:2012walk}
	K.~B. Gustafson and P.~Ricci, ``{L{\'e}vy walk description of suprathermal ion
		transport}'', {\em Physics of Plasmas} {\bfseries 19} (2012), no.~3, 2304.
	
	\bibitem[Humphries et~al.(2010)Humphries, Queiroz, Dyer, Pade, Musyl, Schaefer,
	Fuller, Brunnschweiler, Doyle, Houghton, Hays, Jones, Noble, Wearmouth,
	Southall, and Sims]{Humphries:2010cd}
	N.~E. Humphries, N.~Queiroz, J.~R.~M. Dyer, N.~G. Pade, M.~K. Musyl, K.~M.
	Schaefer, D.~W. Fuller, J.~M. Brunnschweiler, T.~K. Doyle, J.~D.~R. Houghton,
	G.~C. Hays, C.~S. Jones, L.~R. Noble, V.~J. Wearmouth, E.~J. Southall, and
	D.~W. Sims, ``{Environmental context explains L{\'e}vy and Brownian movement
		patterns of marine predators.}'', {\em Nature} {\bfseries 465} (2010),
	no.~7301, 1066--1069.
	
	\bibitem[Jacobs and Sluckin(2014)]{Jacobs:2015gf}
	G.~S. Jacobs and T.~J. Sluckin, ``{Long-range dispersal, stochasticity and the
		broken accelerating wave of advance.}'', {\em Theoretical population biology}
	{\bfseries 100C} (2014) 39--55.
	
	\bibitem[Fogleman et~al.(2001)Fogleman, Fawcett, and Solomon]{Fogleman:2001cc}
	M.~Fogleman, M.~Fawcett, and T.~Solomon, ``{Lagrangian chaos and correlated
		L{\'e}vy flights in a non-Beltrami flow: Transient versus long-term
		transport}'', {\em Physical Review E} {\bfseries 63} (2001), no.~2, 020101.
	
	\bibitem[Del-Castillo-Negrete(2009)]{DelCastilloNegrete:2009dc}
	D.~Del-Castillo-Negrete, ``{Truncation effects in superdiffusive front
		propagation with L{\'e}vy flights}'', {\em Physical Review E} {\bfseries 79}
	(2009), no.~3, 031120.
	
	\bibitem[Peyrard et~al.(2008)Peyrard, Dieckmann, and Franc]{Peyrard:2008kk}
	N.~Peyrard, U.~Dieckmann, and A.~Franc, ``{Long-range correlations improve
		understanding of the influence of network structure on contact dynamics.}'',
	{\em Theoretical population biology} {\bfseries 73} (2008), no.~3, 383--394.
	
	\bibitem[Trapman(2007)]{Trapman:2007ij}
	P.~Trapman, ``{On analytical approaches to epidemics on networks.}'', {\em
		Theoretical population biology} {\bfseries 71} (2007), no.~2, 160--173.
	
	\bibitem[Balcan et~al.(2009)Balcan, Colizza, Goncalves, Hu, Ramasco, and
	Vespignani]{Balcan22122009}
	D.~Balcan, V.~Colizza, B.~Goncalves, H.~Hu, J.~J. Ramasco, and A.~Vespignani,
	``Multiscale mobility networks and the spatial spreading of infectious
	diseases'', {\em Proceedings of the National Academy of Sciences} {\bfseries
		106} (2009), no.~51, 21484--21489.
	
	
	\bibitem[Strano et~al.(2015)Strano, Shai, Dobson, and
	Barthelemy]{Strano20150651}
	E.~Strano, S.~Shai, S.~Dobson, and M.~Barthelemy, ``Multiplex networks in
	metropolitan areas: generic features and local effects'', {\em Journal of The
		Royal Society Interface} {\bfseries 12} (2015), no.~111.
	
	\bibitem[Brockmann and Helbing(2013)]{Brockmann:2013}
	D.~Brockmann and D.~Helbing, ``The hidden geometry of complex, network-driven
	contagion phenomena'', {\em Science} {\bfseries 342} (2013), no.~6164,
	1337--1342.
	
	\bibitem[Keeling(2005)]{Keeling:2005hk}
	M.~Keeling, ``{The implications of network structure for epidemic dynamics.}'',
	{\em Theoretical population biology} {\bfseries 67} (2005), no.~1, 1--8.
	
	\bibitem[Leventhal et~al.(2015)Leventhal, Hill, Nowak, and
	Bonhoeffer]{Leventhal:2015fh}
	G.~E. Leventhal, A.~L. Hill, M.~A. Nowak, and S.~Bonhoeffer, ``{Evolution and
		emergence of infectious diseases in theoretical and real-world networks}'',
	{\em Nature communications} {\bfseries 6} (2015) 6101.
	
	\bibitem[Bayati(2013)]{Bayati:2013}
	B.~S. Bayati, ``Fractional diffusion-reaction stochastic simulations'', {\em
		The Journal of Chemical Physics} {\bfseries 138} (2013), no.~10, 104117.
	
	\bibitem[del Castillo-Negrete et~al.(2003)del Castillo-Negrete, Carreras, and
	Lynch]{CastilloNegrete:2003}
	D.~del Castillo-Negrete, B.~A. Carreras, and V.~E. Lynch, ``Front dynamics in
	reaction-diffusion systems with levy flights: A fractional diffusion
	approach'', {\em Phys. Rev. Lett.} {\bfseries 91} (2003), no.~1, 018302--.
	
	\bibitem[Stollenwerk and Pedro~Boto(2009)]{Stollenwerk2009}
	N.~Stollenwerk and J.~Pedro~Boto, ``Reaction-superdiffusion systems in
	epidemiology, an application of fractional calculus'', {\em AIP Conference
		Proceedings} {\bfseries 1168} (2009) 1548--1551.
	
	\bibitem[Hanert et~al.(2011)Hanert, Schumacher, and
	Deleersnijder]{Hanert:2011cm}
	E.~Hanert, E.~Schumacher, and E.~Deleersnijder, ``Front dynamics in
	fractional-order epidemic models'', {\em Journal of Theoretical Biology}
	{\bfseries 279} (2011), no.~1, 9--16.
	
	\bibitem[Hanert(2012)]{Hanert:2012ka}
	E.~Hanert, ``{Front dynamics in a two-species competition model driven by
		L{\'e}vy flights.}'', {\em Journal of theoretical biology} {\bfseries 300}
	(2012) 134--142.
	
	\bibitem[Draief and Ganesh(2010)]{Draief:2010ct}
	M.~Draief and A.~Ganesh, ``{A random walk model for infection on graphs: spread
		of epidemics {\&} rumours with mobile agents}'', {\em Discrete Event Dynamic
		Systems} {\bfseries 21} (2010), no.~1, 41--61.
	
	\bibitem[Grosche et~al.(2007)Grosche, Rothlauf, and Heinzl]{Grosche:2007fn}
	T.~Grosche, F.~Rothlauf, and A.~Heinzl, ``{Gravity models for airline passenger
		volume estimation}'', {\em Journal of Air Transport Management} {\bfseries
		13} (2007), no.~4, 175--183.
	
	\bibitem[Hanert and Piret(2014)]{Hanert:2014}
	E.~Hanert and C.~Piret, ``A chebyshev pseudospectral method to solve the
	space-time tempered fractional diffusion equation'', {\em SIAM Journal on
		Scientific Computing} {\bfseries 36} (2014), no.~4, A1797--A1812.
	
	\bibitem[Mainardi et~al.(2007)Mainardi, Paradisi, and
	Gorenflo]{Mainardi:2007wc}
	F.~Mainardi, P.~Paradisi, and R.~Gorenflo, ``{Probability distributions
		generated by fractional diffusion equations}'', {\em arXiv.org}, 2007
	
	\bibitem[Kampen(2007)]{Kampen:2007}
	N.~V. Kampen, ``Stochastic processes in physics and chemistry'', North Holland,
	3rd~ed., 2007.
	
	\bibitem[Bayati et~al.(2011)Bayati, Chatelain, and Koumoutsakos]{Bayati:2011}
	B.~Bayati, P.~Chatelain, and P.~Koumoutsakos, ``Adaptive mesh refinement for
	stochastic reaction-diffusion processes'', {\em J. Comp. Phys.} {\bfseries
		230} (2011), no.~1, 13--26.
	
	\bibitem[Jahnke and Altintan(2010)]{Jahnke:2010}
	T.~Jahnke and D.~Altintan, ``Efficient simulation of discrete stochastic
	reaction systems with a splitting method'', {\em BIT} {\bfseries 50(4)} Jan
	(2010) 797--822.
	
	\bibitem[Mainardi et~al.(1997)Mainardi, Paradisi, and Gorenflo]{Mainardi:1997}
	F.~Mainardi, P.~Paradisi, and R.~Gorenflo, ``Probability distributions
	generated by fractional diffusion equations'', {\em International Workshop on
		Econophysics}, 1997.
	
	\bibitem[Gorenflo et~al.(2002{\natexlab{a}})Gorenflo, Mainardi, Moretti,
	Pagnini, and Paradisi]{Gorenflo:2002}
	R.~Gorenflo, F.~Mainardi, D.~Moretti, G.~Pagnini, and P.~Paradisi, ``Discrete
	random walk models for space-time fractional diffusion'', {\em Chemical
		Physics} {\bfseries 284} (2002){\natexlab{a}} 521--541.
	
	\bibitem[Gorenflo et~al.(2002{\natexlab{b}})Gorenflo, Fabritiis, and
	Mainardi]{Gorenflo:2002b}
	R.~Gorenflo, G.~D. Fabritiis, and F.~Mainardi, ``Discrete random walk models
	for symmetric levy-feller diffusion processes'', {\em Physica A} {\bfseries
		269} (2002){\natexlab{b}} 79--89.
	
	\bibitem[Gorenflo and Mainardi(1998)]{Gorenflo:1998}
	R.~Gorenflo and F.~Mainardi, ``Fractional calculus and stable probability
	distributions'', {\em Arch. Mech.} {\bfseries 50} (1998) 1--10.
	
	\bibitem[Gillespie(1976)]{Gillespie:1976}
	D.~Gillespie, ``A general method for numerically simulating the stochastic time
	evolution of coupled chemical reactions'', {\em J. Comput. Phys.} {\bfseries
		22} (1976), no.~4, 403--434.
	
	\bibitem[Gillespie(1977)]{Gillespie:1977}
	D.~Gillespie, ``Exact stochastic simulation of coupled chemical reactions'',
	{\em J. Phys. Chem.} {\bfseries 81} (1977), no.~25, 2340--2361.
	
	\bibitem[Elf and Ehrenberg(2004)]{Elf:2004}
	J.~Elf and M.~Ehrenberg, ``Spontaneous seperation of bi-stable biochemical
	systems into spatial domains of opposite phases'', {\em Syst. Biol.}
	{\bfseries 1} (2004), no.~2, 230--235.
	
	\bibitem[Bernstein(2005)]{Bernstein:2005}
	D.~Bernstein, ``Simulating mesoscopic reaction-diffusion systems using the
	gillespie algorithm'', {\em Phys. Rev. E} {\bfseries 71} (2005), no.~4,
	041103.
	
	\bibitem[Lampoudi et~al.(2009)Lampoudi, Gillespie, and Petzold]{Lampoudi:2009}
	S.~Lampoudi, D.~T. Gillespie, and L.~R. Petzold, ``The multinomial simulation
	algorithm for discrete stochastic simulation of reaction-diffusion systems'',
	{\em J. Chem. Phys.} {\bfseries 130} Mar (2009) 094104.
	
	\bibitem[Drawert et~al.(2010)Drawert, Lawson, Petzold, and
	Khammash]{Drawert:2010}
	B.~Drawert, M.~J. Lawson, L.~Petzold, and M.~Khammash, ``The diffusive finite
	state projection algorithm for efficient simulation of the stochastic
	reaction-diffusion master equation'', {\em J. Chem. Phys.} {\bfseries 132}
	Feb (2010) 074101.
	
	\bibitem[Hellander and Lotstedt(2007)]{Hellander:2007}
	A.~Hellander and P.~Lotstedt, ``Hybrid method for the chemical master
	equation'', {\em J. Comput. Phys.} {\bfseries 227} (2007), no.~1, 100--122.
	
	\bibitem[Hellander et~al.(2012)Hellander, Hellander, and
	Lotstedt]{Hellander:2012}
	A.~Hellander, S.~Hellander, and P.~Lotstedt, ``Coupled mesoscopic and
	microscopic simulation of stochastic reaction-diffusion processes in mixed
	dimensions'', {\em Multiscale Modeling \& Simulation} {\bfseries 10} (2012),
	no.~2, 585--611.
	
	\bibitem[Gillespie(2001)]{Gillespie:2001}
	D.~Gillespie, ``Approximate accelerated stochastic simulation of chemically
	reacting systems'', {\em J. Chem. Phys.} {\bfseries 115} (2001), no.~4,
	1716--1733.
	
	\bibitem[Auger et~al.(2006)Auger, Chatelain, and Koumoutsakos]{Auger:2006}
	A.~Auger, P.~Chatelain, and P.~Koumoutsakos, ``R-leaping: Accelerating the
	stochastic simulation algorithm by reaction leaps'', {\em J. Chem. Phys.}
	{\bfseries 125} Aug (2006) 084103.
	
	\bibitem[Gillespie(2007)]{Gillespie:2007}
	D.~Gillespie, ``Stochastic simulation of chemical kinetics'', {\em Annu. Rev.
		Phys. Chem.} {\bfseries 58} (2007) 35--55.
	
	\bibitem[Ginsberg et~al.(2009)Ginsberg, Mohebbi, Patel, Brammer, Smolinski, and
	Brilliant]{Ginsberg2009}
	J.~Ginsberg, M.~H. Mohebbi, R.~S. Patel, L.~Brammer, M.~S. Smolinski, and
	L.~Brilliant, ``Detecting influenza epidemics using search engine query
	data'', {\em Nature} {\bfseries 457} (2009), no.~7232, 1012--1014.
	
	\bibitem[Lazer et~al.(2014)Lazer, Kennedy, King, and Vespignani]{Lazer14032014}
	D.~Lazer, R.~Kennedy, G.~King, and A.~Vespignani, ``The parable of google flu:
	Traps in big data analysis'', {\em Science} {\bfseries 343} (2014), no.~6176,
	1203--1205,
	
	\bibitem[Coburn et~al.(2009)Coburn, Wagner, and Blower]{Coburn2009}
	B.~J. Coburn, B.~G. Wagner, and S.~Blower, ``Modeling influenza epidemics and
	pandemics: insights into the future of swine flu (h1n1)'', {\em BMC Medicine}
	{\bfseries 7} (2009) 30--30.
	
	\bibitem[Bayati and Eckhoff(2012)]{Bayati2012}
	B.~S. Bayati and P.~A. Eckhoff, ``Influence of high-order nonlinear
	fluctuations in the multivariate susceptible-infectious-recovered master
	equation'', {\em Phys. Rev. E} {\bfseries 86} (2012), no.~6, 062103--.
	
\end{thebibliography}
\begingroup\raggedright\endgroup

\end{document}